\documentclass[twocolumn, amssymb, nobibnotes, aps, pre, superscriptaddress, floatfix, epsfig, graphics]{revtex4}

\usepackage{graphicx}
\usepackage{amsmath}
\usepackage{dcolumn}% Align table columns on decimal point
\usepackage{bm}% bold math
\usepackage{accents}

\begin{document}

\title{Theory of supercooled water}
\author{Jacobo Troncoso}
\email{jacobotc@uvigo.es}
\author{Claudio A. Cerdeiri\~na}
\email{calvarez@uvigo.es}
\affiliation{Instituto de F\'isica, Computaci\'on e Ciencia Aeroespacial da Universidade de
  Vigo and Unidad MSMN Asociada al CSIC por el IQF Blas Cabrera, Ourense 32004, Spain} 

\date{\today}

\begin{abstract}

We introduce a model of water contemplating true supercooled-liquid states that, as such, are metastable with respect to the crystalline-solid ones. Its numerical solutions reproduce from Speedy-Angell's stability-limit picture to Poole et al.'s metastable liquid--liquid criticality. The real existence of a liquid--liquid critical point is found to gain significant theoretical plausibility from a hitherto unnoticed analogy with the ``inversion thermodynamics'' of Joule-Thomson gas throttling process and, by extension, with the equivalent phenomenology exhibited by black holes.

\noindent
\end{abstract}

\maketitle

\noindent

\emph{Introduction}---Thermodynamics sharply distinguishes water from common simple substances such as argon or methane. At moderate pressure $p$, the volume per particle $v$ of ice is unusually higher than that of its coexisting liquid. Likewise unusual is that liquid's $v$ exhibits a minimum as a function of temperature $T$ along isobars ---at $\sim 277$ K at ambient $p$. This generates a locus of states, usually referred to as the TMD locus \cite{TMD}, that separates regions in which the isobaric thermal expansivity $\alpha_p$ takes on values of opposite sign. The pattern of behavior becomes even more intriguing when water is cooled below the freezing point while maintained in a metastable state as a supercooled liquid, with the magnitude of $\alpha_p$ and other thermodynamic response functions such as the isothermal compressibility $\kappa_T$ and the specific heat $c_p$ exhibiting sharp rises \cite{speedyangell,oguni,sorensen,caupin,nilssonkT,nilssoncp}.

The past fifty years have seen a flourishing discussion on physical pictures accommodating water's unusual thermodynamics. The first of them posits that response functions increase because a stability limit is approached as $T$ is lowered \cite{speedyangell,retracing-spinodal,speedypoints}. A second possibility revealed by molecular simulation is that it is a second, liquid--liquid critical point what is approached \cite{poolestanley,palmer,advances,pgdscience,wail,lltransition2,mbpol}. In addition, response-function enhancement can be alternatively explained without any appeal to a thermodynamic singularity \cite{singularityfree,singularityfreeII}. State-of-the-art experiments are providing increasing evidence supporting the second-critical-point picture \cite{kringle,nilsson,nilsson-2,nilsson-3}.

The problem can be studied theoretically via analytically tractable models based on statistical mechanics, which are cheap computationally and allow to gain insights from the perspective of physical congruence and plausibility. Thus, on properly characterizing water's hydrogen bonding, an increasing number of such simple models has provided theoretical support to the second-critical-point scenario \cite{poolemodel,debenedetti,mercedesbenz,franzese,ciach,franzeseII,cacstanley,cagey,isingparadigm,caupinanisimov,coronasfranzese,entropy,dill}. A few models reproducing stability-limit and singularity-free scenarios also exist \cite{speedypoints,singularityfree,singularityfreeII}. Nevertheless, all these models lack a description of the crystalline phase and the freezing transition, implying that the metastable nature of the liquid phase is overlooked. Such an incomplete characterization of supercooled water prevents a full theoretical understanding.

\begin{figure}
    \centering
    \includegraphics[width=1.0\linewidth]{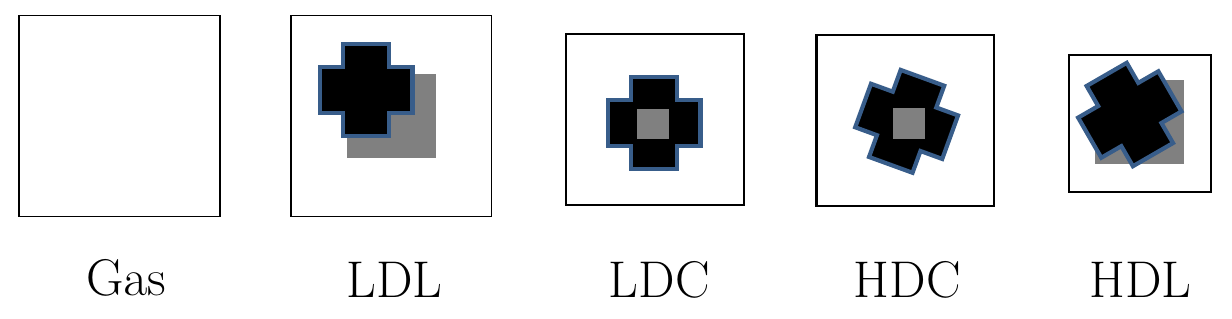}
    \caption{Two-dimensional illustration of individual cell states of our five-state Ising model. Shaded in grey are the free volumes explored by particles (crosses). Associated with each cell state is a phase: see text.}
    \label{fig:placeholder}
\end{figure}

\begin{figure*}
    \centering
    \includegraphics[width=1.0\linewidth]{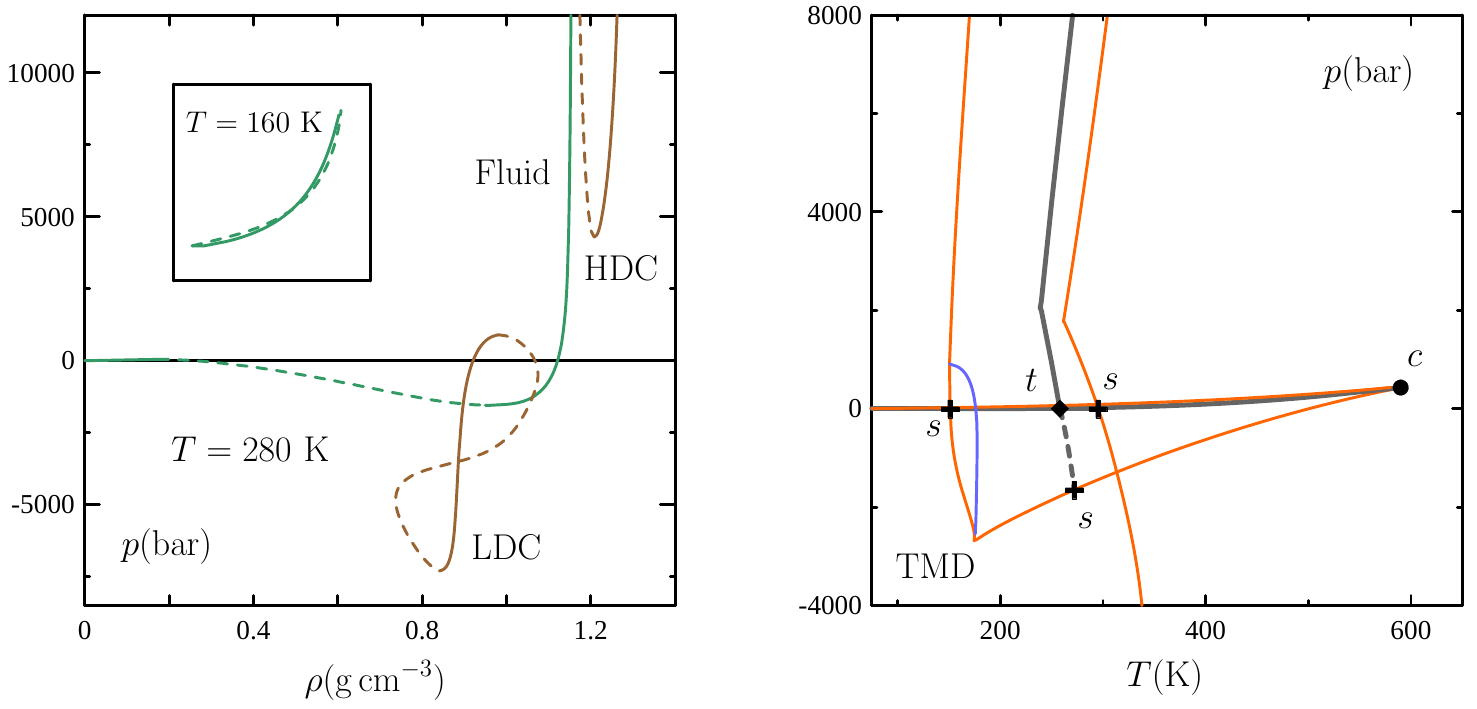}
    \caption{Numerical results for our five-state model with parameter setting ``1.'' (Left) Pressure $p$ versus density $\rho$ along an isotherm of temperature $T$. Three independent branches display stable and metastable states for fluid, LDC, and HDC phases, dashed lines being unstable sections. The inset shows the fluid branch at a lower $T$. (Right) Corresponding phase diagram in the pressure-temperature $p$-$T$ plane. Plotted are binodals (thick, grey), spinodals (thin, orange), and the locus of temperatures of maximum density along isobars TMD (thin, blue). Special points are critical $c$, triple $t$, and $s$ points. Continuations of crystal spinodals beyond their crossing point are not drawn for clarity of presentation; likewise for the freezing binodals.}
    \label{fig:placeholder}
\end{figure*}

The five-state Ising model introduced here provides the desirable complete approach to the problem \cite{beg}. In particular, on combining our recent prototypes for crystalline solids, freezing, and water's fluid phases \cite{cacstanley,isingparadigm,entropy,prefreezing,premetastability}, we manage to devise a variant of the class capable to reproduce from stability-limit to second-critical-point scenarios with the liquid rigorously metastable with respect to the crystal. A mean-field treatment is pertinent since its unambiguous description of metastable states enables a proper characterization of supercooled water \cite{mean-field}. Our discussion places special emphasis on the theoretical congruence of the various physical pictures. In this connection, the second-critical-point scenario gains significant plausibility from a striking analogy with the thermodynamics of Joule-Thomson throttling process of gases that, by extension, establishes a link with the thermodynamics of black holes \cite{hawking,chamblin,kastor,dolan,kubiznak,blackhole}.

\emph{Model}---We need to describe crystal, liquid, and gas phases. This demands going beyond the most basic Ising variants with two states for individual lattice sites, which are in general suited to just two phases. The physical objects associated with lattice sites ---spins for the case of a ferromagnet--- are in our model ``cells'' filling out space, their state being characterized by attributes such as their volume or the number of particles they contain. In this connection, recent work has shown that four cell states are needed to properly describe the full phase behavior and metastable states of a simple substance \cite{prefreezing,premetastability}. It is then natural to expect that a complex substance like water requires more states. On restricting ourselves to the gas, two liquid phases, and just two ice polymorphs, we find that five states are needed \cite{plethora}.

Figure 1 illustrates the molecular details featured by our five cell states. We distinguish between vacant cells and cells occupied by a nonspherical particle with orientational degrees of freedom whose center of mass explores a free volume. A given state is then defined by cell's occupancy, volume, free volume, and number of particle's orientations, all of which regarded in general distinct for distinct states \cite{mef}. The fluctuating nature of cell volumes and number of particle's orientations allows to characterize phases with distinct densities and hydrogen-bond network efficiencies \cite{cacstanley,isingparadigm,entropy}. In turn, fluctuating free volumes distinguish the positional order of crystalline solids from disordered configurations \cite{prefreezing,premetastability}. A one-to-one correspondence between cell states and phases follows from Fig. 1. The gas phase is rich in vacant cells. Larger cells containing particles with a relative high degree of orientational selectivity distinguishes a low-density liquid LDL with ``ice-like order'' from a normal, high-density one HDL. In addition, when appropriate volumetric changes are coupled to nearest-neighbor pairs, cells containing particles with positional order lead to a high-density crystal HDC and a low-density one LDC mimicking ice Ih. The microscopic picture is completed by energetic couplings between pairs of like and unlike cells, which drive the transitions between the five phases.

The populations of the distinct cell states define four order parameters whose overall values are central to our mean-field approach. This gives rise to a generalization of the single-order-parameter theory developed by van der Waals. The resulting $pvT$ relation enables, upon parameter setting, the calculation of thermodynamic properties. The phase behavior is obtained from a standard analysis of isotherms in terms of binodal and spinodal points respectively determining coexistence and stability limits.

\begin{figure}
    \centering
    \includegraphics[width=1.0\linewidth]{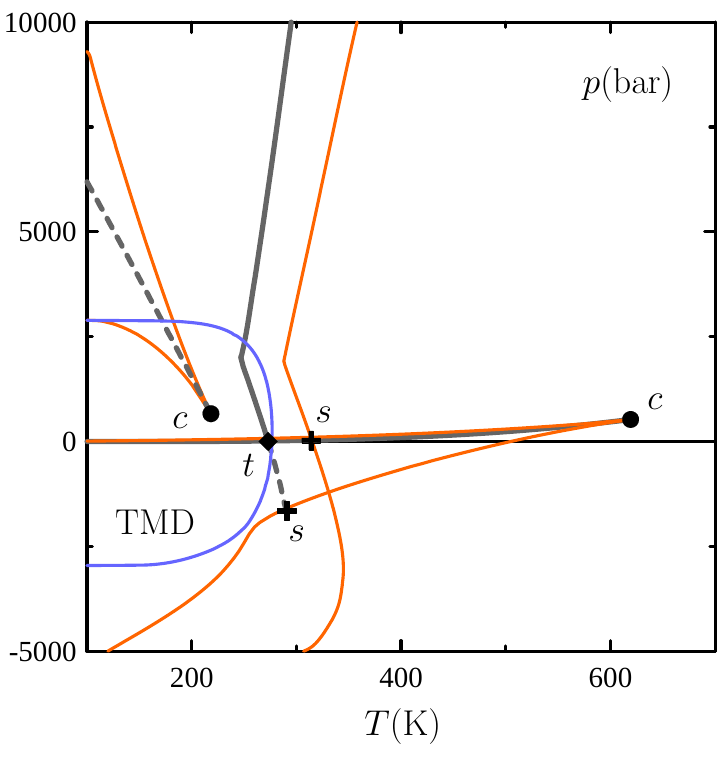}
    \caption{Numerical results for our five-state model with parameter setting ``2.'' Plotted in the $p$-$T$ plane are binodals (thick, grey), spinodals (thin, orange), and the locus of temperatures of maximum density along isobars TMD (thin, blue). Special points are critical $c$, triple $t$, and $s$ points. Continuations of crystal spinodals beyond their crossing point are not drawn for clarity of presentation; likewise for the freezing binodals.}
    \label{fig:placeholder}
\end{figure}

\emph{Speedy-Angell's stability-limit scenario}---Figure 2 shows, for a model's parameter setting, the phase diagram in the $p$-$T$ plane together with $p$ as a function of $\rho$ along a $T = 280$ K isotherm. The isotherm identifies three main branches corresponding to fluid, HDC, and LDC phases. The fluid branch displays the usual van der Waals loop leading to the standard gas--liquid binodal curve terminating at a critical point $c$. It furthermore leads to two spinodal curves, one of them determining a limit for gas overpressurizing and the other for liquid superheating and stretching ($p<0$). The HDC branch displays a spinodal point at its minimum, while the LDC one is bounded above and below by two spinodal points at its maximum and minimum. For each crystal branch, there is a $p$ value at which the chemical potential takes on the same value than that for the fluid branch at the same $p$. This obviously corresponds to phase equilibrium and leads to the two freezing binodal curves of the $p$-$T$ diagram, which correspond to HDC--liquid and LDC--liquid coexistence. Associated with such binodals are two spinodal curves for crystal superheating and a low-$T$ one for liquid supercooling. The LDC--liquid binodal intersects the gas--liquid one at the standard triple point $t$ of crystal--liquid--gas coexistence. At that $t$ point do metastable continuations of binodal loci develop until they reach endpoints $s$ at spinodals ---this was noted originally by Speedy \cite{speedy-pgd} and later confirmed by others \cite{speedypoints,caupinanisimov,premetastability}.

We have checked that response functions diverge on approaching the supercooled liquid spinodal \cite{spinodaldiverging}. Therefore, in essence, our phase diagram in Fig. 2 realizes the picture early envisioned by Speedy and Angell for the experimentally observed low-$T$ rises of response functions \cite{speedyangell,retracing-spinodal}. The supercooled liquid spinodal branch meets the superheating-stretching one at a cusp point. This is a prediction by Stillinger \cite{stillinger1995}, who revised the original hypothesis of a smooth $p(T)$ minimum \cite{retracing-spinodal}. Furthermore, the smoothness of our entire supercooled liquid spinodal also contrasts with the form with a cusp conjectured originally \cite{retracing-spinodal}. Certainly, the shape of the branch replicates the unusual isentropic $T(p)$ behavior of liquid water and its two extrema bound the TMD locus in accord with a thermodynamic constraint imposing that the TMD can only meet a spinodal at a point where $(dp/dT)_{\rm spinodal}$ changes sign \cite{retracing-spinodal}.

\emph{Poole et al.'s metastable liquid--liquid criticality}---An alternative model's variant characterized by an alternative parameter setting leads to the phase diagram in Fig. 3. The supercooled liquid spinodal is now absent. It is superseded by a binodal corresponding to coexistence between HDL and LDL phases and terminating at a second $c$ point. This is the second-critical-point scenario put forth by Poole et al. \cite{poolestanley}. The whole transition is metastable with respect to the crystalline solid phase, which certainly corresponds to the absolute minimum of the Gibbs free energy. Our model thus provides a theoretical realization of metastable coexistence and criticality of two truly supercooled liquid phases, as supported by experiment and simulation \cite{pgdscience,wail,lltransition2,mbpol,nilsson,nilsson-3}.

The liquid--liquid transition in Fig. 3 retains the form known from models contemplating fluid phases alone \cite{poolemodel,debenedetti,mercedesbenz,franzese,ciach,franzeseII,cacstanley,cagey,isingparadigm,caupinanisimov,coronasfranzese,entropy,dill}. Hence, incorporation of the crystal merely makes the transition to become metastable. Furthermore, the so-called ``critical-point-free scenario'' \cite{speedypoints} is obtained upon modification of the value of a single model parameter. As per quantitative performance, note that we have parametrized our model so as to meet the experimental 277 K value for the temperature of maximum density at ambient $p$. This expediency yields $T_c\approx 218$ K for the HDL--LDL critical temperature, which may be regarded reasonably consistent with recent estimates from experiment and simulation placing it around 210 K \cite{mbpol,coronasfranzese,nilsson-3,vega}.

\emph{Inversion thermodynamics}---Classical thermodynamics itself provides significant plausibility to the second-critical-point scenario. This statement originates from Fig. 4. Its upper left panel illustrates water's unusual thermodynamics, namely, isobaric $v(T)$ curves exhibit a minimum that entails a $p(T)$ minimum along isochores and defines a locus, the TMD, separating regions with inverted signs for the values of the thermal pressure coefficient $\gamma_v\equiv (\partial p/\partial T)_v$. The upper right panel shows an analogous pattern of behavior for argon, namely, isothermal $h(p)$ curves ---with $h$ the enthalpy per particle--- exhibit a minimum that entails a $T(p)$ maximum along isenthalpics and defines a locus separating regions with inverted signs for the values of the Joule-Thomson coefficient $\mu_{\rm JT}\equiv (\partial T/\partial p)_h$. Inversion of the sign of $\mu_{\rm JT}$ is the signature of the effect uncovered by Joule and Thomson as early as 1852 in connection with the throttling process of gases. Such an inversion of $\mu_{\rm JT}$ for every fluid is analogous to that of $\gamma_v$ for liquid water at low $T$. Certainly, we face in both cases a pattern of ``inversion thermodynamics.''

\begin{figure}
    \centering
    \includegraphics[width=1.0\linewidth]{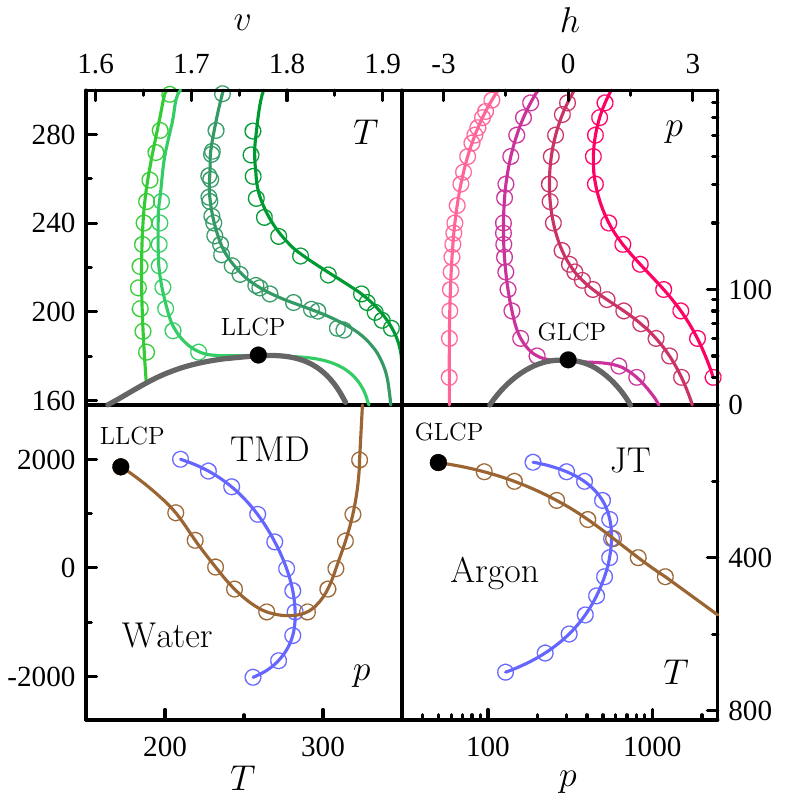}
    \caption{(Left) Molecular simulation results for the TIP4P/2005 water force field \cite{pgdscience,vegadata,sumisekino,comprehensive}. (Right)  Experimental data for argon \cite{nist}. The upper panels display isobars in the temperature-volume per particle $T$-$v$ plane and isotherms in the pressure-enthalpy per particle $p$-$h$ plane, with binodal curves in grey. The lower panels show the corresponding TMD and Joule-Thomson JT inversion loci (blue) together with the loci of isobaric extrema of the isothermal compressibility $\kappa_T$ and isothermal extrema of the isobaric heat capacity per particle $c_p$ (brown) respectively emanating from the water's liquid--liquid critical point LLCP and argon's gas--liquid critical point GLCP. Units are: $p{\rm (bar)}$, $T{\rm (K)}$, $v{\rm (10^{-5}\,m^3\,mol^{-1})}$, and $h{\rm (10^{-3}\,J\,mol^{-1})}$.}
    \label{fig:placeholder}
\end{figure}

\begin{table}
\caption{Analogies in inversion thermodynamics.}
\label{Table1}
\renewcommand
{\tabcolsep}{3pt}
\begin{tabular}{cccc}
\hline \hline 
 & Fluid & Low-$T$ water & Black hole$^a$ \\
\hline
Critical point & Gas-liquid & Liquid--liquid & Small--large \\
Density$^b$ & $h$ & $v$ & $M$ \\
Variable field$^b$ & $p$ & $T$ & $-{\Lambda\over 8\pi}$ \\ 
Constant field$^b$ & $T$ & $p$ & $T_{\rm H}$ \\
Locus & Joule-Thomson & TMD & Inversion \\
\hline \hline
\end{tabular}
\footnotetext{$^a$M stands for the mass of the black hole, $\Lambda$ for the cosmological constant, and $T_{\rm H}$ for the Hawking temperature. $^b$The terminology ``fields and densities'' follows up on \cite{griffithswheeler}.}
\end{table}

Clearly, $v$ is the analog of $h$, while $T$ and $p$ play interchanged roles. Accordingly, $\alpha_p\equiv (\partial v/\partial T)_p/v$ and $\kappa_T\equiv -(\partial v/\partial p)_T/v$ are the respective analogs of $\eta_T\equiv (\partial h/\partial p)_T$ and $c_p\equiv (\partial h/\partial T)_p$. The $\alpha_p$-$\eta_T$ analogy is implicit in the upper panels of Fig. 4, which make it clear that derivatives of isobaric $v(T)$ curves correspond to derivatives of isothermal $h(p)$ curves. In turn, the lower panels of Fig. 4 certify the $\kappa_T$-$c_p$ analogy: the locus of isobaric $\kappa_T(T)$ extrema intersects the TMD locus at the point at which $(dT/dp)_{\rm TMD}=0$, just as the locus of isothermal $c_p(p)$ extrema intersects the JT inversion locus at the point at which $(dp/dT)_{\rm JT}=0$ \cite{singularityfree,singularityfreeII,rowlinson}.

Also common to both patterns is that part of them is a critical point, liquid--liquid for low-$T$ water and gas--liquid for argon. Significantly, the universal character of the gas--liquid transition, its termination at criticality, and its concomitant Joule-Thomson inversion curve jointly indicate that an inversion curve is usually associated with a critical point. This renders plausibility for water's experimentally observed TMD locus to be the prelude of a second critical point, as most recent experimental observations confirm \cite{nilsson-3}. The converse situation may be reasonably regarded as the implausible one and, from this perspective, water's behavior is unusual mainly because a pattern of inversion thermodynamics associated with criticality manifests twice.

Table I summarizes the analogies between low-$T$ water and fluids. In addition, it includes correspondence with black hole thermodynamics originating from relatively recent work illuminating that charged anti-de Sitter black holes also exhibit Joule-Thomson-like phenomena \cite{blackhole}. Again, the problem includes a critical point \cite{chamblin,kastor,dolan,kubiznak}. Hence there is a suggestion that inversion thermodynamics may show up for other systems displaying critical points. To the extent that this is the case, a unified description of a phenomenology common to areas of condensed matter and other branches of physics is called for.

\emph{Outlook}---Extension of our present model entails incorporating the glass to its coarse-grained description of crystal, liquid, and gas. A first step forward would be a mean-field theory contemplating crystalline, fluid, and vitreous states for a simple substance. A challenging chapter would then be its generalization to water, for which glass phenomenology is intriguing and believed to be crucially relevant to the liquid--liquid transition \cite{mishimastanley,pgd,loertingsciortino}. This faces extra hurdles since the current mean-field theory of glasses omits the crystal and remains to be fully developed beyond the simplest cases \cite{parisi}.

\emph{Acknowledgments}---Support from the Spanish Ministry of Science, Innovation, and Universities under grant no. PID2023-147148NB-I00 is greatly acknowledged.

\newpage

%
%
%\begin{figure} [!h]
%\includegraphics[scale=0.90]{504730JCP_fig1.eps}% Here is how to import EPS art
%\vspace{2 in}  \caption{\label{fig:fig1} }
%\end{figure}

%\newpage

%\begin{figure} [!h]
%\includegraphics[scale=0.90]{504730JCP_fig2.eps}% Here is how to import EPS art
%\vspace{2 in}  \caption{\label{fig:fig2} }
%\end{figure}

%\newpage
%\begin{figure}[!h]
%\includegraphics[scale=0.80]{504730JCP_fig3.eps}% Here is how to import EPS art
%\vspace{2 in}  \caption{\label{fig:fig3} }
%\end{figure}

%\newpage

%\begin{figure}[!h]
%\includegraphics[scale=0.75]{504730JCP_fig4.eps}% Here is how to import EPS art
%\vspace{2 in}  \caption{\label{fig:fig4} }
%\end{figure}

%\newpage

%\begin{figure}[!h]
%\includegraphics[scale=0.75]{504730JCP_fig4.eps}% Here is how to import EPS art
%\vspace{2 in}  \caption{\label{fig:fig4} }
%\end{figure}

%\newpage

%\begin{figure} [!h]
%\includegraphics[scale=0.75]{504730JCP_fig5.eps}% Here is how to import EPS art
%\vspace{2 in}  \caption{\label{fig:fig5} }
%\end{figure}

%\newpage

%\begin{figure} [!h]
%\includegraphics[scale=0.75]{504730JCP_fig7.eps}% Here is how to import EPS art
%\vspace{2 in}  \caption{\label{fig:fig7} }
%\end{figure}

%\newpage

%\include{504730JCP_fig_captions}


\begin{thebibliography}{120}
\bibitem{TMD} Note that a $v(T)$ minimum corresponds to a $\rho(T)$ maximum, with $\rho\equiv v^{-1}$ the number density. This implies that the locus of temperatures of maximum density TMD is also that of temperatures of minimum volume per particle. We use throughout the paper the acronym TMD since it is the most popular one.
\bibitem{plethora} No account for the plethora of ice phases is provided. Taking into account that underlying them is a number of distinct crystalline lattices, one should not realistically expect to characterize the full range of ice polymorphs from a mean-field calculation that, as such, is insensitive to lattice structure. In any event, a crude description of ice phases is enough for our present purposes, which put the focus on supercooled water and thus basically render the crystal the role of an spectator phase with respect to which the liquid is metastable.
\end{thebibliography}
\end{document}